\documentclass{aa}

\usepackage{graphics}
\usepackage{longtable}
\usepackage{amssymb}

\begin{document}

%\thesaurus{06     % A&A Section 6: Form. struct. and evolut. of stars
%              (08.09.2;  % Stars: individual: Procyon
%               03.20.7;  % Techniques: radial velocities 
%               08.15.1)} % Stars: oscillations (including pulsations)

\title{Solar--like oscillations in Procyon A\thanks{Based on observations collected with the \textsc{Coralie} echelle spectrograph on the 1.2--m Euler Swiss telescope at La Silla Observatory, ESO Chile.}}

\author{P.~Eggenberger \and F.~Carrier \and F.~Bouchy \and A.~Blecha}
   
   \institute{Observatoire de Gen\`eve, 51 Chemin des Maillettes, CH--1290 Sauverny, Switzerland}

   \offprints{P. Eggenberger\\
   \email{Patrick.Eggenberger@obs.unige.ch}}

\date{Received / Accepted }
\titlerunning{Solar--like oscillations in Procyon A}

\abstract{
The F5 subgiant Procyon A ($\alpha$ CMi, HR~2943) was observed with the \textsc{Coralie} fiber--fed echelle spectrograph on the 1.2--m Swiss telescope at La Silla 
in February 1999. The resulting 908~high--accuracy radial velocities exhibit a mean noise level in the amplitude spectrum
of 0.11~m\,s$^{-1}$ at high frequency. These measurements show significant excess in the power spectrum between 0.6-1.6 mHz with 0.60~m\,s$^{-1}$ peak amplitude. An average
large spacing of 55.5 $\mu$Hz has been determined and twenty-three individual frequencies have been identified. 
\keywords{Stars: oscillations -- Stars: individual: $\alpha$ Canis Minoris
             -- Techniques: radial velocities}
}
   \maketitle

\section{Introduction}
The success of helioseismology encourages corresponding investigations on other stars. The measurement of frequencies and amplitudes of p--mode oscillations for solar--like stars provides an insight into the internal structure and is currently the most powerful constraint on the theory of stellar evolution
(see Guenther \& Demarque \cite{guenther}). 

Solar--like oscillation modes generate periodic motions of the stellar
surface with periods in the range of 3--30 minutes but with extremely small
amplitudes. Essentially, two methods exist to detect such a motion:
photometry and Doppler spectroscopy. In photometry, the oscillation 
amplitudes of solar--like stars
are within 2--30~ppm, while they are in the range of
10--150~cm\,s$^{-1}$ in radial velocity measurements.
Photometric measurements made from the ground are strongly limited by
scintillation noise. To reach the needed accuracy requires
observations made from space.
In contrast, Doppler ground--based measurements have recently shown their
ability to detect oscillation modes in solar--like stars.
Since 1999, solar--like oscillations have been detected in a growing 
list of main sequence and subgiant stars (see the review by 
Bouchy \& Carrier \cite{bc03}).

Procyon is the first northern hemisphere candidate for the 
search for p--mode oscillations. Several groups have made thorough attempts to
detect the signature of oscillation modes on this bright F5 subgiant star.
The first convincing result was obtained by Martic et al. (\cite{martic}) 
with observations conducted with the \textsc{Elodie} spectrograph. These Doppler 
observations led to the detection of p--modes in the 0.6--1.6 mHz frequency range 
with amplitudes 2 times solar and a most likely frequency 
spacing equal to 55 $\mu$Hz, but no individual frequency was identified.
On the theoretical side, several stellar evolution models
of Procyon have been computed predicting a large spacing between 50 and 60~$\mu$Hz (Barban et al. \cite{ba99}, Chaboyer et al. \cite{ch99}, 
Di Mauro \& Christensen--Dalsgaard \cite{di01}, Provost et al. \cite{pr02} and Kervella et al. \cite{ke04}).

In this paper, we report Doppler observations 
of Procyon~A made with the \textsc{Coralie} spectrograph, well known for 
the characterization of p--modes on $\alpha$ Cen A (Bouchy \& Carrier \cite{bc01}, 
\cite{bc02}). 
These new measurements, made independently with a different calibration method
than the one used by Martic et al. (\cite{martic}) on \textsc{Elodie} (simultaneous thorium method instead of a
tunable Fabry-Perot illuminated by a white source), confirm the detection of p--modes and enable
the identification of twenty-three individual mode frequencies.

The data reduction and the acoustic spectrum analysis of Procyon~A are presented in Sect.~2 and 3, respectively.
The conclusion is given in Sect.~4.

\begin{figure}[thb]
\resizebox{\hsize}{!}{\includegraphics{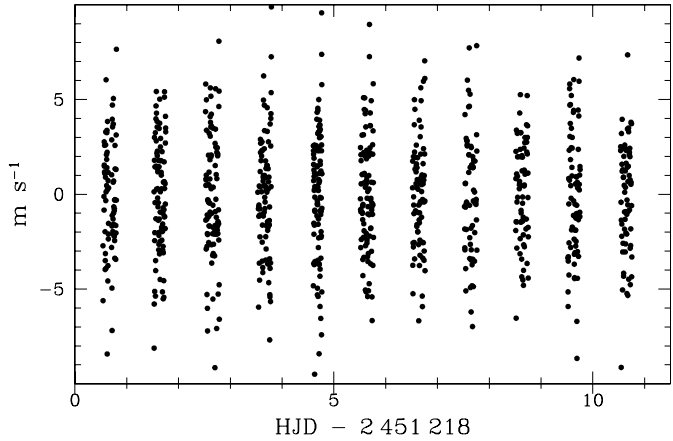}}
\caption[]{Radial velocities of Procyon relative to a reference spectrum taken during each night (the best spectrum of the night). The dispersion (which includes the noise and the oscillations) reaches 3.1\,m\,s$^{-1}$.}
\label{Figure vr_temps}
\end{figure}

\section{Observations and data reduction}

Procyon was observed over a campaign of eleven nights (8--18~February~1999) with \textsc{Coralie}, the high--resolution fiber--fed echelle spectrograph 
mounted on the 1.2--m Swiss telescope at La Silla (ESO, Chile).

A description of the spectrograph and the data reduction process is presented 
in Bouchy \& Carrier (\cite{bc01}) and Bouchy et al. (\cite{bouchy}). The exposure times were 
60\,s with 110\,s dead--times  in between. In total, 908 optical spectra were 
obtained with a signal--to--noise (S/N) ratio in the range of 200 to 300 at 550 nm. 
For each night, radial velocities were computed relative to the
highest signal--to--noise ratio optical spectrum obtained in the middle of the
night (when the target had the highest elevation). 
The radial velocity measurements are shown in Fig.~\ref{Figure vr_temps} and their distribution and 
dispersion are listed in Table~\ref{tab:alphacmi}.
The dispersion of these measurements reaches 3.1\,m\,s$^{-1}$.

\begin{table}
\caption{Distribution and dispersion of Doppler measurements.}
\begin{center}
\begin{tabular}{clll}
\hline
\hline
\multicolumn{1}{c}{Date} & \multicolumn{1}{c}{Nb spectra} & \multicolumn{1}{c}{Nb hours} &
 \multicolumn{1}{c}{$\sigma$[m s$^{-1}$]}  \\ \hline
1999/02/08 & 76 & 6.22 & 2.83 \\
1999/02/09 & 89 & 5.43 & 2.92 \\
1999/02/10 & 98 & 6.36 & 2.97 \\
1999/02/11 & 93 & 6.30 & 2.95 \\
1999/02/12 & 97 & 4.00 & 3.45 \\
1999/02/13 &  95& 6.04 & 3.03 \\
1999/02/14 &  79& 5.78 & 2.59 \\
1999/02/15 &  68& 5.72 & 3.17  \\
1999/02/16 & 68 & 5.84 & 2.56 \\
1999/02/17 & 75 & 5.65 & 3.32 \\
1999/02/18 &  70& 5.19 & 2.61 \\ 
\hline
\label{tab:alphacmi}
\end{tabular}
\end{center}
\end{table}

\section{Stellar power spectra analysis}
\label{sectps}
\begin{figure}[thb]
\resizebox{\hsize}{!}{\includegraphics{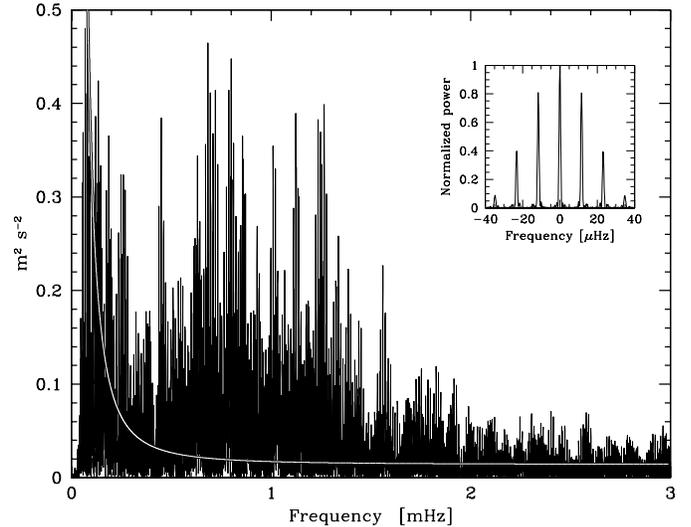}}
\caption[]{Power spectrum of the radial velocity measurements of Procyon with the observational window response. The white line indicates the noise
in the power spectrum.}
\label{Figure power}
\end{figure}

In order to compute the power spectrum of the velocity time series, we use the Lomb--Scargle modified 
algorithm (Lomb \cite{lomb}, Scargle \cite{scargle}) giving directly a power in m$^{2}$\,s$^{-2}$ without the need for
correction by other normalization factors. The time scale gives a resolution of 1.13~$\mu$Hz. Fig.~\ref{Figure power} 
shows this periodogram. Typically for such a power spectrum, the noise has two components:
\begin{itemize}
\item At high frequencies it is flat, indicative
of the Poisson statistics of photon noise. The mean white noise level $\sigma_{\mathrm{pow}}$ calculated between 2 and 2.5~mHz 
is 0.015\,m$^2$\,s$^{-2}$, namely 0.11\,m\,s$^{-1}$ in amplitude. With 908 measurements, this high frequency noise 
corresponds to $\sigma_{RV}\,=\,\sqrt{N \sigma_{\mathrm{pow}} /4 }\,=\,1.85$~m\,s$^{-1}$.
\item Towards the lowest frequencies, the power should scale inversely with frequency squared as expected for instrumental instabilities. 
However, the computation of the radial velocities introduces a high pass filter. Indeed, the radial velocities were computed relative to one
reference for each night and the average radial velocities of the night fixed to zero (see Sect.~2). This results in an 
attenuation of the very low frequencies which can be seen on Fig.~\ref{Figure power}.
\end{itemize}
The power spectrum presents an excess in the range 0.6--1.6~mHz. 
The combined noise has a value decreasing from 0.027 to 0.016\,m$^2$\,s$^{-2}$ in the above--mentioned interval (see Fig.~\ref{Figure power}).
The noise has been determined by fitting a function of the type 1/frequency$^2$ without considering the attenuated values at very low
frequencies. Note that the filtering induced by the radial velocities computation does not influence the frequency of the peaks in the 
range 0.6--1.6~mHz, but could slightly change their amplitudes.
 The amplitude of the strongest peaks reaches 0.60~m\,s$^{-1}$, corresponding to a signal to noise of 5 (in the amplitude spectrum).
This amplitude is estimated as the height of the peaks in the power spectrum with a quadratic subtraction of the mean noise level. 
To investigate if the peaks are due to noise or are p--modes, we have conducted simulations in which we analyzed noise spectra
containing no signal. 
For this purpose, a velocity time series is build, using the observational time sampling
and radial velocities randomly drawn by assuming a Gaussian noise (Monte--Carlo
simulations). The simulated noise does not need to include a spectral dependance, since the threshold determination is
made relative to $\sigma$ (see below). Indeed, the spectral dependance is already included in the observed $\sigma$ (noise),
which varies with frequency in the power spectrum (see Fig.~\ref{Figure power}).
The amplitude spectrum of this series is then calculated and peaks with
amplitude greater than 3,
4 and 5\,$\sigma$ are counted; note that a peak and its aliases are only counted once. 
The whole procedure is repeated 1000 times to ensure the
stability of the results. In this way, we find that the number of peaks due to noise with an amplitude
larger than 3\,$\sigma$ varies between 0 and 12 in the range 0.6--1.6~mHz, with a mean value of 2.6 and a standard deviation
of 3.4.
For 4\,$\sigma$, the number of peaks due to noise varies between 0 and 2 in the range 0.6--1.6~mHz, 
with a mean value of 0.0 and a standard deviation of 0.3. 
No peaks due to noise are expected with an amplitude larger than 
5\,$\sigma$. These very convincing results show that the power excess is due to p--modes.

\subsection{Search for a comb--like pattern}
\label{combsect}
In solar--like stars, p--mode oscillations are expected to produce a
characteristic comb--like structure in the power spectrum with mode
frequencies
$\nu_{n,\ell}$ reasonably well approximated by the simplified asymptotic
relation (Tassoul \cite{tassoul80}):
\begin{eqnarray}
\label{eq1}
\nu_{n,\ell} & \approx &
\Delta\nu(n+\frac{\ell}{2}+\epsilon)-\ell(\ell+1)\delta\nu_{02}/6
\end{eqnarray}
with $\Delta\nu\,=\,\langle \nu_{n,\ell}-\nu_{n-1,\ell}\rangle$ and \\
$\delta\nu_{02}\,=\,\langle \nu_{n,0}-\nu_{n-1,2}\rangle$\,.\\

The two quantum numbers $n$ and $\ell$ correspond to the radial
order and the angular degree of the modes, respectively. $\Delta\nu$ and
$\delta\nu_{02}$
are the large and small spacing. For stars whose disk is
not resolved, only the lowest--degree modes ($\ell\,\leq\,3$) can be detected.
In the case of stellar rotation, the degeneracy of the modes is lifted and p--modes need to be
characterized by a third quantum number $m$ called the azimuthal order:
\begin{eqnarray}
\label{eq2}
\nu_{n,\ell,m} & \approx & \nu_{n,\ell,0}+m\tilde{\Omega}_{n\ell m}/2\pi\,,
\end{eqnarray}
with $-\ell\,\leq\,m\,\leq\,\ell$ and $\tilde{\Omega}_{n\ell m}$ the averaged angular velocity over the whole volume of 
the star.

\begin{figure}[thb]
\resizebox{\hsize}{!}{\includegraphics{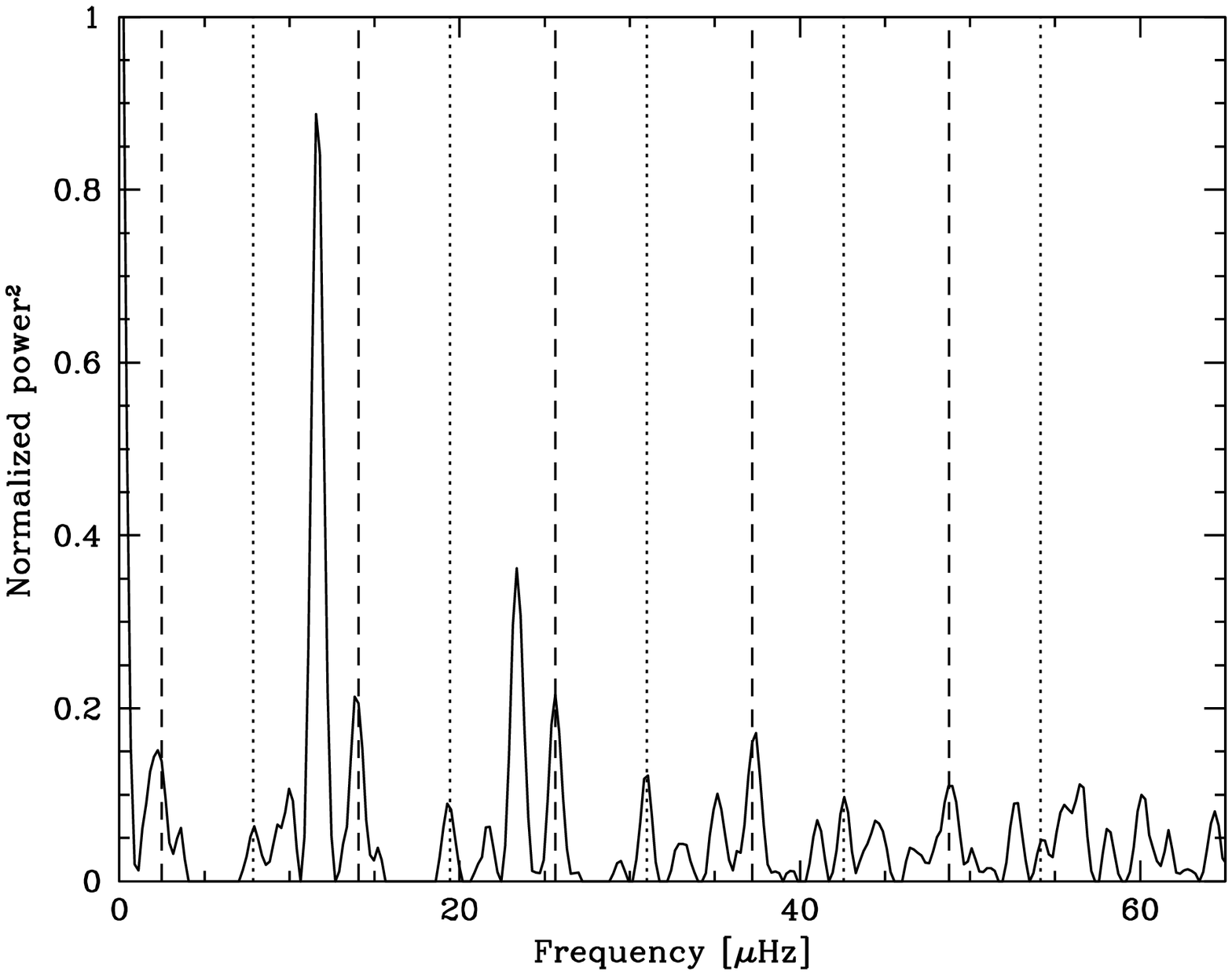}}
\caption[]{Autocorrelation of the power spectrum of Procyon with a threshold of 0.25~m$^{2}$\,s$^{-2}$. The dashed lines correspond to 25.5~$\mu$Hz and its aliases, while the dotted ones indicate the value of 31~$\mu$Hz and its aliases (see Sect.~\ref{combsect}).}
\label{autocor}
\end{figure}

One technique, commonly used to search for periodicity in the power
spectrum,
is to compute its autocorrelation.
To reduce the uncertainties due to the noise, only peaks greater than 0.25\,m$^{2}$\,s$^{-2}$ in the power spectrum (corresponding to a S/N of 4
in the amplitude spectrum at 1.6~mHz)
in the frequency range 0.6\,--\,1.6~mHz have been used to compute the autocorrelation shown in Fig.~\ref{autocor}.

The two strongest peaks at 11.5 and 23\,$\mu$Hz correspond to the daily alias. Two other peaks at 25.5 and 31\,$\mu$Hz and their daily aliases can be identified. 
According to stellar evolution models of Procyon (Barban et al. \cite{ba99}, Chaboyer et al. \cite{ch99}, Di Mauro \& Christensen--Dalsgaard \cite{di01}
and Provost et al. \cite{pr02}), the large spacing $\Delta\nu$ is expected to lie between 50 and 60 $\mu$Hz. Fig. 3 exhibits no peak having a significant amplitude in this frequency interval.
It is not surprising that a clear peak corresponding to the value of the averaged large spacing $\Delta\nu$ cannot be identified with an
autocorrelation, since theoretical models predict that $\Delta\nu$ varies sensitively (several $\mu$Hz) with the frequency between 0.6 and 1.6 mHz. Thus, if
only a few consecutive oscillation modes are observed, the different values of the large spacing will not contribute to raising the amplitude of a single peak
in the autocorrelation, but will instead give a succession of small peaks around its averaged value.   

In the case of Procyon, it is much more interesting to look for characteristic spacings between modes with different angular degrees. Indeed,
theoretical models predict that, contrary to the large spacing $\Delta\nu$, these spacings remain approximately constant with the frequency. 
The peak at 25.5 $\mu$Hz can be identified as the spacing between the modes $\ell=1$ and $\ell=0$ with the same radial order and the spacing 
between $\ell=2$ and $\ell=1$, also with the same radial order.    
The peak at 31 $\mu$Hz results from the spacing between the modes ($\ell=0$, $n=n_0$) and ($\ell=1$, $n=n_0-1$), as well as the spacing 
between ($\ell=1$, $n=n_0$) and ($\ell=2$, $n=n_0-1$). The autocorrelation spectrum thus suggests that the large spacing must be close to 
the sum of these two peaks, i.e. close to 56 $\mu$Hz.

\begin{figure}[thb]
\resizebox{\hsize}{!}{\includegraphics{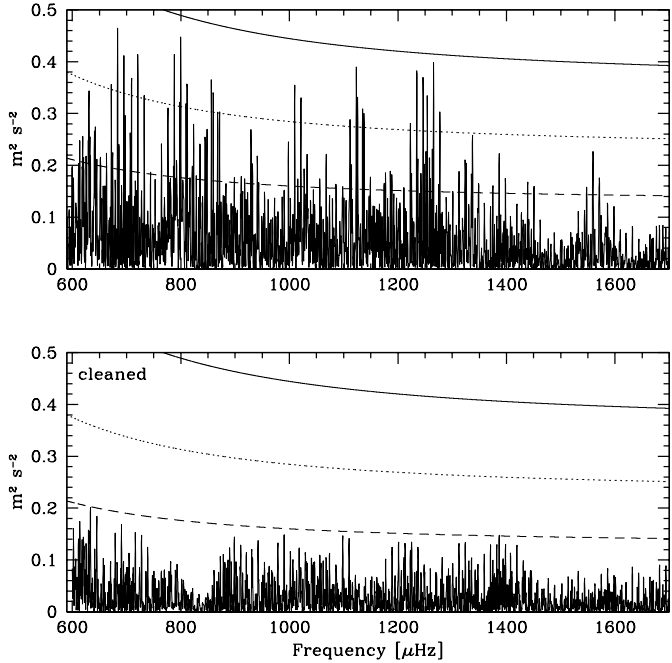}}
\caption[]{{\bf Top:} Original power spectrum for the eleven nights of observation. {\bf Bottom:} Cleaned power spectrum: all peaks listed
in Table~\ref{tab:identif} have been removed. The continuous, dotted and dashed lines indicate an amplitude of 5\,$\sigma$, 
4\,$\sigma$ and 3\,$\sigma$, respectively. Numerous peaks are still present below 3\,$\sigma$, 
since no peaks have been cleaned below this threshold. These peaks can be due to p--mode oscillations and noise or have been
artificially added by the extraction algorithm due to the finite lifetimes of the modes} 
\label{clean}
\end{figure}

\begin{table}
\caption[]{Identification of extracted frequencies. Some frequencies can
be either split $\ell =2$ mode or due to noise. The number of peaks due to noise is in agreement with the
simulations described in Sect.~\ref{sectps}, which predict $2.6 \pm 3.4$ noise peaks.}
\begin{center}
\begin{tabular}{rcc}
\hline
\hline
\multicolumn{1}{c}{Frequency} & Mode ID  & S/N \\
\multicolumn{1}{c}{$[\mu$Hz$]$} &           &     \\
\hline
 $651.5 - 11.57 = 639.9$  & $\ell = 2$  &  3.2  \\
 $630.8 + 11.57 = 642.4$  & $\ell = 0$  &  3.9  \\
     662.7              & noise       &  3.2  \\
 $683.5 + 11.57 = 695.1$  & $\ell = 0$  &  4.7  \\
     720.6              & $\ell = 1$  &  4.5  \\
     797.9              & $\ell = 2$ &  4.1  \\
     799.7              & $\ell = 2$  &  4.8  \\
 $791.8 + 11.57 = 803.4$  & $\ell = 0$  &  3.2  \\
     828.5              & $\ell = 1$  &  3.0  \\
     835.4              & noise       &  3.6  \\
     856.2              & $\ell = 2$  &  4.4  \\
     859.8              & $\ell = 0$  &  4.3  \\
     911.4              & $\ell = 2$  &  3.2  \\
 $929.2 - 11.57 = 917.6$  & $\ell = 0$  &  3.9  \\
 $929.2 + 11.57 = 940.8$  & $\ell = 1$  &  3.9  \\
 $1009.7 - 11.57 = 998.1$ & $\ell = 1$  &  4.5  \\
     1027.1             & $\ell = 0$  &  3.4  \\
 $1123.3 + 11.57 = 1134.9$& $\ell = 2$ &  4.8  \\
     1137.0             & $\ell = 2$  &  4.0  \\
 $1131.1 + 11.57 = 1142.7$& $\ell = 0$  &  3.0  \\
     1192.4             & $\ell = 2$  &  3.0  \\
 $1186.0 + 11.57 = 1197.6$& $\ell = 0$  &  3.4  \\
 $1234.8 + 11.57 = 1246.4$& $\ell = 2$  &  4.9  \\
     1251.8             & $\ell = 0$  &  3.7  \\
 $1265.6 + 11.57 = 1277.2$& $\ell = 1$  &  5.0  \\
     1337.2             & noise       &  4.0  \\
     1439.0             & noise       &  3.3  \\
     1559.5             & $\ell = 1$? &  3.7  \\
\hline
\end{tabular}
\end{center}
\label{tab:identif}
\end{table}

\subsection{Echelle diagram}

\begin{figure*}[thb]
\begin{center}
\resizebox{\hsize}{!}{\includegraphics{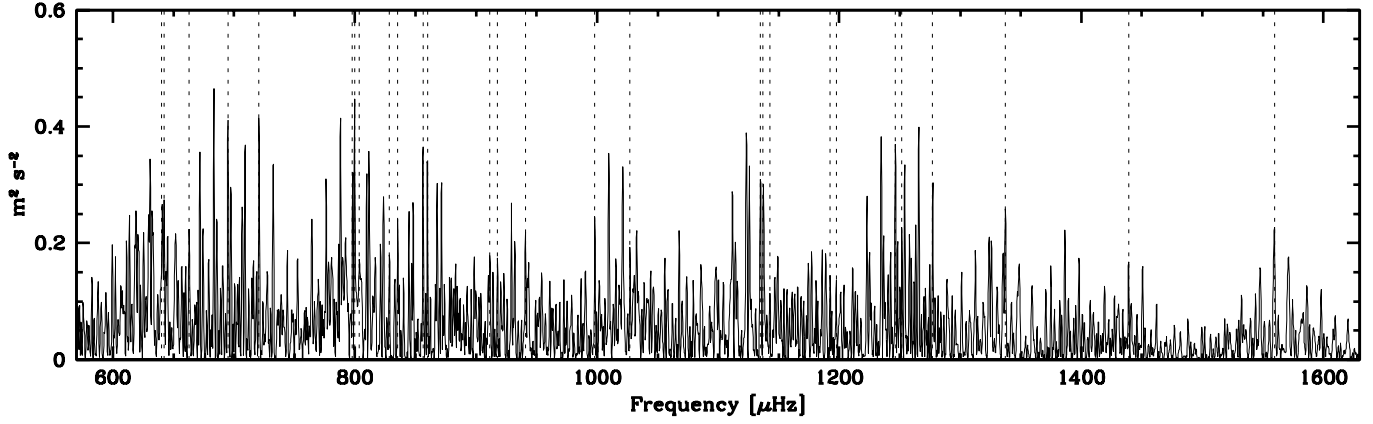}}
\caption[]{Power spectrum of Procyon with the twenty-eight extracted frequencies indicated by dotted lines. The identification of each
extracted frequency is given in Table~\ref{tab:identif}.}
\label{tfiden}
\end{center}
\end{figure*}

To identify the individual frequencies of the modes, we made echelle diagrams corresponding to values of the averaged large spacing around 
56 $\mu$Hz. 

The frequencies were extracted using an iterative algorithm that identifies the highest peak between
0.6 and 1.6\,$\mu$Hz and subtracts
it from the time series. Note that because of the stochastic nature of solar--like oscillations, a timestring
of radial velocities cannot be expected to be perfectly reproduced by a sum of sinusoidal terms. Therefore, using
an iterative clean algorithm to extract the frequencies can add additional peaks with small amplitudes due to the finite
lifetimes of the modes that we do not know. Nevertheless, the iterative algorithm ensures that one peak and its aliases
with an amplitude above a given threshold is only extracted once.
To avoid extracting artificial peaks with small amplitudes added by the iterative
algorithm, the choice of this threshold is important. In the case of Procyon, 
we iterated the process until all peaks with an amplitude higher than 3\,$\sigma$ in the amplitude
spectrum were removed (see Fig. \ref{clean}). Note that peaks with amplitudes below the 3\,$\sigma$ threshold were 
not considered not only to avoid extracting artificial peaks added by the iterative
algorithm, but also because they were too strongly influenced by noise and interactions between noise and daily aliases.
This threshold, which ensures that the selected peaks have
only a small
chance to be due to noise, gave a total of twenty-eight frequencies (see Table~\ref{tab:identif}).
Because of the daily alias of 11.57 $\mu$Hz introduced by the monosite observations (see the
observational
window response in Fig.~\ref{Figure power}), we cannot know {\it a priori} whether the frequency
selected by the algorithm is the right one or an alias. Thus, we considered
that the frequencies could be shifted by $\pm 11.57$ $\mu$Hz. We then made echelle diagrams
for different large spacings between 50 and 60 $\mu$Hz until each frequency could be identified as
an $\ell=0$, $\ell=1$, $\ell=2$ mode or attributed to noise (see Table~\ref{tab:identif}). 
In this way, we found an averaged large spacing of 55.5 $\mu$Hz.   

At 799 and 1136\,$\mu$Hz, either the $\ell=2$ modes are split by the rotation or the second peaks are due to
the noise. 
Allende Prieto et al. (\cite{al02}) recently estimated $v \sin i =3.16 \pm 0.50$\,km\,s$^{-1}$ but mentioned that the correct value
is probably close to 2.7\,km\,s$^{-1}$, given that the value of 3.16\,km\,s$^{-1}$ may be slightly overestimated  
due to the finite numerical resolution of their convection simulation.
Using $v \sin i =2.7$\,km\,s$^{-1}$ and
$i=31.1\pm 0.6^{\circ}$ (Girard et al. \cite{gi00}), the rotational splitting expected for Procyon A is about 0.6~$\mu$Hz. 
For $\ell=2$ modes, this splitting can thus results in differences between two modes as high as 2.4~$\mu$Hz, 
which is the frequency separation between $m=-2$ and $m=+2$ modes.
Of course, the quality of our data is not good enough to unambiguously determine this rotational splitting.
The possible $\ell=1$ peak 
at 1559.5\,$\mu$Hz is rejected, as it is far from the other modes with a doubtful identification.
Moreover, the frequency of 1027.1\,$\mu$Hz is difficult to identify; we identify it as a $\ell=0$ mode,
but it could also be a $\ell=2$ mode.

\begin{figure}[thb]
\resizebox{\hsize}{!}{\includegraphics{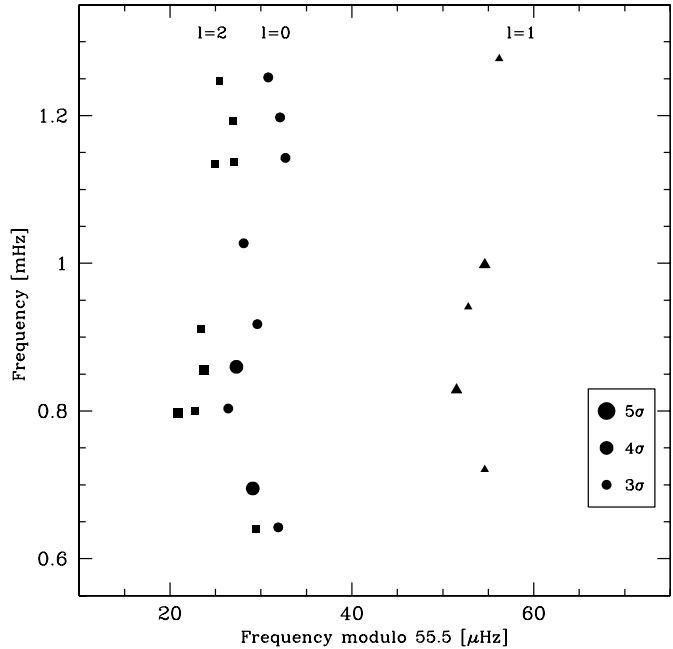}}
\caption[]{Echelle diagram for the frequencies
  listed in Table~\ref{tab:freq}. Dots, triangles and squares correspond respectively to modes identified as
  $\ell=0$, $\ell=1$ and $\ell=2$. The size of the symbols is proportional to the signal-to-noise ratio of the mode.}
\label{echobs}
\end{figure}

The echelle diagram showing the twenty-three identified modes is shown in Fig.~\ref{echobs}. 
The frequencies of the modes are given in Table~\ref{tab:freq}.

\begin{table}
\caption{Mode frequencies (in $\mu$Hz) of Procyon.}
\begin{center}
\begin{tabular}{ccc}
\hline
\hline
$\ell$ = 0 & $\ell$ = 1 & $\ell$ = 2 \\
\hline
	 &         & 639.9 \\
 642.4   &  	   &       \\
 695.1   & 720.6   &       \\
         & 	   & 797.9 / 799.7 \\
 803.4   & 828.5   & 856.2  \\
 859.8   &	   & 911.4\\
 917.6   & 940.8   & \\
  	 & 998.1   &        \\
 1027.1  & 	   &        \\
         & 	   & 1134.9 / 1137.0 \\
 1142.7  & 	   & 1192.4    \\
 1197.6  & 	   & 1246.4  \\
 1251.8  & 1277.2  &   	    \\
\hline
\end{tabular}\\
\end{center}
\label{tab:freq}
\end{table}

\subsection{Amplitudes of the modes}

Concerning the amplitudes of the modes, theoretical computations predict oscillation amplitudes for a 1.5 $M_{\odot}$ star like Procyon~A 
between 1 and 2 m\,s$^{-1}$, with mode lifetimes of the order of a few days (Houdek et al. \cite{ho99}). The observed amplitudes of 0.6 m\,s$^{-1}$ are 
then lower than expected. This disagreement can be partly explained by the lifetimes of the modes. Indeed, the oscillation modes have finite lifetimes, 
because they are continuously damped. Thus, if the star is observed for a time longer than the lifetimes of the modes, the signal is weakened due to 
the damping of the modes and to their re--excitation with a random phase.

\section{Conclusion}

The radial velocity measurements of Procyon~A, obtained over 11 nights, show a significant excess in the power spectrum between 0.6--1.6~mHz, 
centered around 0.8~mHz, with a peak amplitude of 0.60~m\,s$^{-1}$,
revealing solar--like oscillations with a large spacing of 55.5 $\mu$Hz.
Our results confirm the values of the large spacing and of the amplitudes of the modes found by Martic et al. (\cite{martic}).

Moreover, we presented the identification of twenty-three individual frequencies. Note that this identification is seriously complicated
by the presence of daily aliases. 
Interactions between real peaks and aliases can
slightly shift the frequencies of the modes. This is particularly true for the $\ell=0$ and $\ell=1$ modes which are separated by 25.5~$\mu$Hz: 
the interactions between the second alias of the radial modes (separated from the real peak by 23.14~$\mu$Hz)
and the $\ell=1$ modes can slightly shift the observed $\ell=1$ frequencies, although the different 
peaks are resolved. Inversely, the second alias of the $\ell=1$ modes can influence the observed frequencies of the radial modes. 
Nevertheless, we think that this identification is reliable, since it is the only one that can explain all the highest peaks
in the power spectrum. 

To obtain more accurate parameters and to thoroughly test the physics of the models, it is not
only important to obtain better data, but also to eliminate the mode identification ambiguity
due to the aliases. The space mission \textsc{Most} (Matthews \cite{matthews}) launched 
on 30 June 2003 will observe Procyon~A during a whole month and should therefore provide severe constraints on this star.

\begin{acknowledgements}
This work was partly supported by the Swiss National Science Foundation. 
\end{acknowledgements}

\end{document}